# Speeding up high-throughput characterization of materials libraries by active learning: autonomous electrical resistance measurements

Felix Thelen[1], Lars Banko[1], Rico Zehl[1], Sabrina Baha[1], Alfred Ludwig[1,*]

[1]Chair for Materials Discovery and Interfaces, Institute for Materials, Ruhr University Bochum, Universitätsstraße 150, 44801 Bochum, Germany

*Corresponding author: alfred.ludwig@rub.de

**Abstract**
High-throughput experimentation enables efficient search space exploration for the discovery and optimization of new materials. However, large search spaces of, e.g., compositionally complex materials, require decreasing characterization times significantly. Here, an autonomous measurement algorithm was developed, which leverages active learning based on a Gaussian process model capable of iteratively scanning a materials library based on the highest uncertainty. The algorithm is applied to a four-point probe electrical resistance measurement device, frequently used to obtain indications for regions of interest in materials libraries. Ten materials libraries with different complexities of composition and property trends are analyzed to validate the model. By stopping the process before the entire library is characterized and predicting the remaining measurement areas, the measurement efficiency can be improved drastically. As robustness is essential for autonomous measurements, intrinsic outlier handling is built into the model and a dynamic stopping criterion based on the mean predicted covariance is proposed. A measurement time reduction of about 70-90% was observed while still ensuring an accuracy of above 90%.

**Introduction**
The emerging of complex materials such as high entropy alloys (HEA) or compositionally complex solid solutions (CCSS) results in an immense multidimensional search space, making the use of efficient research methods and strategies mandatory [1]. One approach of dealing with this complexity is combinatorial materials science and high-throughput experimentation, which involves synthesizing a large number of materials in parallel and performing rapid automated characterization of a variety of materials properties [2].

High-throughput experiments usually consist of three main stages, starting with the combinatorial fabrication of hundreds of well-defined chemical compositions in the form of thin-film materials libraries [3]. These can either have a continuous compositional gradient, e.g., generated by co-deposition magnetron sputtering [4], or can be ordered discretely, i.e., by inkjet printing techniques [5], [6]. An example of a co-sputtered materials library is shown in Figure 1. After fabrication, the libraries are characterized by multiple techniques ideally in parallel or by automated serial methods. These include, first, identification of the chemical compositions and their crystallographic structure, e.g., by energy dispersive X-ray analysis (EDX) and X-ray diffraction (XRD) respectively. Second, functional properties are investigated based on the use cases of the fabricated materials, which include for example electrical resistance or band gap measurements [1]. Thereby, most high-throughput characterization instruments consist of an automated positioning system which moves a sensor system over the materials library. After characterization, the large amounts of data generated along these steps are then used to plan follow-up experiments. Many characterization techniques remain, however, rather time-consuming compared to the synthesis process, e.g., performing XRD measurements for hundreds of measurement areas on a single library can take 12-14 hours [7].

Especially for the last stage, the application of machine learning and data mining under the paradigm of materials informatics [8], [9] has contributed significantly to navigate, explore, and exploit the high-dimensional materials search space more efficiently. In order to decrease the necessary time for high-throughput characterization, active learning together with Gaussian process (GP) regression can be leveraged to autonomously determine materials properties across libraries more efficiently. Instead of measuring all, typically hundreds of measurement areas of a library consecutively with fixed coordinates, the algorithm decides the measurement sequence by building and updating a Gaussian process model during the procedure. Once the model's prediction is accurate enough, the process can be terminated, decreasing the total measurement time drastically: related work [10] indicates a 10-fold time reduction. An essential factor for autonomous characterization is the robustness of a model as it needs to be applicable to a wide variety of materials and measurement procedures can be affected by systematic measurement errors.

To investigate the possibilities and limitations of this approach, the algorithm is tested on a custom-built high-throughput test-stand [11] measuring the electrical resistance of materials libraries using the four-point probe method. The electrical resistivity in alloys is dependent on the crystal structure and is further influenced by all defects in the materials as electrons are scattered at lattice defects like voids, impurities, dislocations, and grain boundaries [7]. Therefore, mapping of the resistance of a library can indicate different phase zones/regions and their boundaries [1], [11] and is thus a useful descriptor for finding areas of interest.

Ten libraries comprising a variety of metallic materials systems fabricated with different methods such as co- and multilayer-sputtering, were measured and analyzed to validate the performance of the developed algorithm.

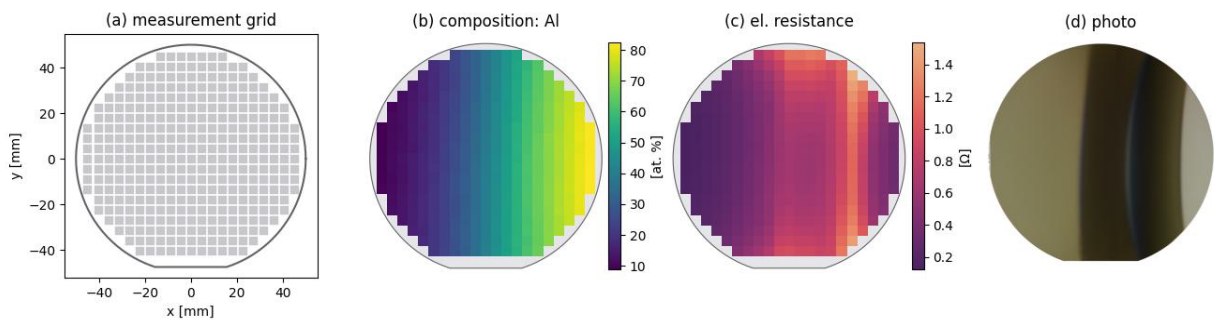

**Figure 1:** Example of a thin-film materials library used for the exploration and exploitation of materials search spaces. Here a 100 mm diameter Ni-Al library fabricated by magnetron co-sputtering is shown as a basic example with simple compositional gradients. Al was deposited from the right, Ni from the left. (a) shows the applied measurement grid with 342 measurement areas. The composition map (b) obtained by EDX shows a linear gradient between the two elements. The electrical resistance (c) varies along the compositional gradient and shows similarities to the visual information shown in (d).

**Methods**

*Four-Point Probe Measurement*

Thin film resistance is measured by the four-point probe method [12], [13]. The probe consists of four spring contacts (Feinmetall F238, $d < 0.3\ mm$) with an equidistant spacing of about $0.5\ mm$. Due to the attachment to a computer-controlled stage movable in x, y and z-direction (UHL F9S-3-M), the contact of vertical as well as the horizontal positioning of the pins with respect to the sample can be adjusted on the micrometer level. By using a source meter (Keithley 2400), a direct current $I_0$ is induced in the two outer contact pins while the voltage $V$ is measured at the two inner pins. In order to accommodate for varying pin orientations resulting from the touchdown of the probe, each measurement area on the library is measured three times and during each touch down, ten measurements are conducted to decrease measurement noise. The components are controlled via a custom application implemented in Python running on an Intel® Core™ i7 8GB RAM Windows PC [10].

*Active Learning in Materials Discovery*

In the majority of machine learning approaches, a learner is treated as a passive recipient of the data by providing whole training datasets at once [14]. In contrast, active learning differs from that approach by allowing an algorithm to choose the data from which it learns, resulting in a higher performance with less training effort [15]. In an active learning process, a surrogate model is iteratively choosing from a pool of unknown training data via a query algorithm. The selected instances are then passed to an oracle (e.g., a human annotator or a measurement system), which assigns the instances with a label. After labelling, the model is updated with the query result [16]. Depending on the intended purpose of the algorithm, a variety of learning models are available, e.g., support vector machines, naïve Bayes, decision trees or neural networks [14]. In the regression setting, a Gaussian process is used most often due to its flexibility and ability of uncertainty quantification independently from the actual observations [17], [18]. A Gaussian process is a generalization of the multivariate Gaussian probability distribution, which describes the relation of n-random variables depending on a mean vector $\boldsymbol{\mu}$ and a covariance matrix $\boldsymbol{\Sigma}$. In stochastic processes like the Gaussian process, every point of a function $(\boldsymbol{x}_i|f(\boldsymbol{x}_i))$ is treated as a single random variable. These points can then be approximated by adjusting the mean vector and covariance matrix of the Gaussian process, described in this setting as the mean function $m(\boldsymbol{x})$ and covariance function $k(\boldsymbol{x}, \boldsymbol{x}')$ [19]. In terms of the mean function, $m(\boldsymbol{x}) = 0$ is most often assumed, as the data can be standardized, and the Gaussian process is generally flexible enough to model the mean sufficiently well [20]. The covariance function consists of a kernel function, which returns the similarity of two random variables and therefore controls the function's shape. A variety of different kernels exists, which each have their own set of hyperparameters. The kernel needs to be selected depending on the use case, the most widely used being the squared exponential kernel and the Matérn kernels [18], [21], [22].

This learning approach is especially useful in scenarios in which labels are expensive or time-consuming to generate. Therefore, active learning fits the conditions of materials discovery with its mostly elaborate measurement techniques [23]. Example of applications of active learning for materials discovery can be found in [24]–[26]. Closely related to active learning is Bayesian optimization, which is in comparison more frequently applied in the field of materials discovery. In contrast to active learning, instead of the goal of learning an underlying function as efficiently as possible, Bayesian optimization aims to maximize a function globally [18]. As materials discovery most often has the aim to identify materials with optimized properties while reducing the number of experiments, Bayesian optimization is applied frequently in literature [10], [23], [27]–[32].

*Active Learning for Autonomous Measurement Processes*
The measurement algorithm was implemented in Python using the *GPflow* [33] package which is based on *TensorFlow*. Figure 2 shows the structure of the algorithm. Before incorporating any training data, the Gaussian process predicts the same mean and covariance for the entire library. Therefore, the procedure is initialized with nine predefined measurement areas evenly distributed across the library. A too small number of initialization areas can result in divergence of the process, a too large number reduces the achieved efficiency improvement by the algorithm. Automatic relevance determination was used to increase the flexibility of the model. The initialization measurement areas are labelled by the oracle afterwards, which means the resistance is measured by the described setup. The Gaussian process is then trained on the acquired data by adjusting the model's hyperparameters using marginal likelihood optimization. The resulting model is then used to predict the not yet measured areas afterwards, and the next area is selected based on the predicted covariance. This process continues until a stopping criterion is met.

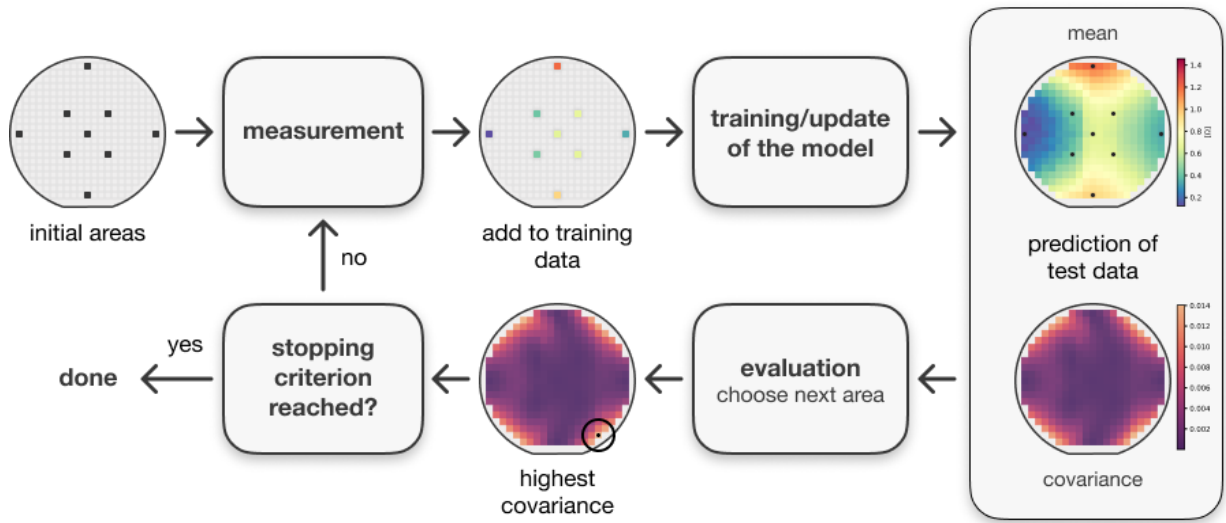

**Figure 2:** Concept of the autonomous measurement process visualized on the example of a Ni-Al library. The Gaussian process (GP) is initialized on nine measurement areas evenly distributed across the library. First, these initial areas are measured and added to the output training data, on which the GP is trained afterwards. Previously obtained composition data is used as an input. After training, the entire library is predicted by the GP and based on the predicted uncertainty, the next area is selected. This procedure is executed until a stopping criterion is met.

**Results and Discussion**
To assess the ability of the algorithms to perform well on a variety of libraries, the ground truth for ten test libraries with different materials systems was fully measured in advance, so that the accuracy of the prediction at each iteration can be determined via the coefficient of determination $R^2$. Table 1 shows an overview of the measured libraries, for further details see supporting information. The test libraries were selected in order to cover different fabrication methods, number of constituents and materials systems. To increase the robustness of the algorithm, modifications to the standard Gaussian process were tested, ranging from the incorporation of the substrate information into the training data and including the measurement variance into the model to standardization of the output training data. Furthermore, a kernel test was done by comparing the performance of the Gaussian process with various kernel functions.

**Table 1:** Materials libraries used to test the autonomous measurements performance.

| Material system | Sputter method | Substrate | Deposition temperature | Annealing temperature | Thickness |
|---|---|---|---|---|---|
| Co-Fe-Mo-Ni-V | co-sputtering | Si+SiO$_2$ | 25 °C | - | ≈ 150 nm |
| Co-Fe-Mo-Ni-W-Cu | co-sputtering | Si+SiO$_2$ | 25 °C | - | ≈ 500 nm |
| Co-Cr-Fe-Mo-Ni | co-sputtering | Si+SiO$_2$ | 25 °C | - | ≈ 150 nm |
| Cr-Fe-Mn-Mo-Ni | co-sputtering | Si+SiO$_2$ | 25 °C | - | ≈ 150 nm |
| Co-Cr-Fe-Mn-Mo | co-sputtering | Si+SiO$_2$ | 25 °C | - | ≈ 150 nm |
| Ni-Al | co-sputtering | Si+SiO$_2$ | 25 °C | - | ≈ 300 nm |
| Co-Cr-W 1 | multilayer | Al$_2$O$_3$ | 150 °C | 900°C | ≈ 300 nm |
| Co-Cr-W 2 | multilayer | Al$_2$O$_3$ | 150 °C | 750°C | ≈ 300 nm |
| Co-Cr-W 3 | multilayer | Al$_2$O$_3$ | 25 °C | 600°C | ≈ 300 nm |
| Co-Cr-Mo | multilayer | Al$_2$O$_3$ | 25 °C | 900°C | ≈ 300 nm |

Figure 3 shows two iterations of the autonomous measurement on the example of the Co-Cr-Fe-Mn-Mo library. After measuring areas for initialization, the algorithm first selects areas at the edge of the library, before concentrating on the inner parts. While parts of the library are still incorrectly predicted after five iterations, the ground truth and the prediction are almost visually identical after 15 iterations.

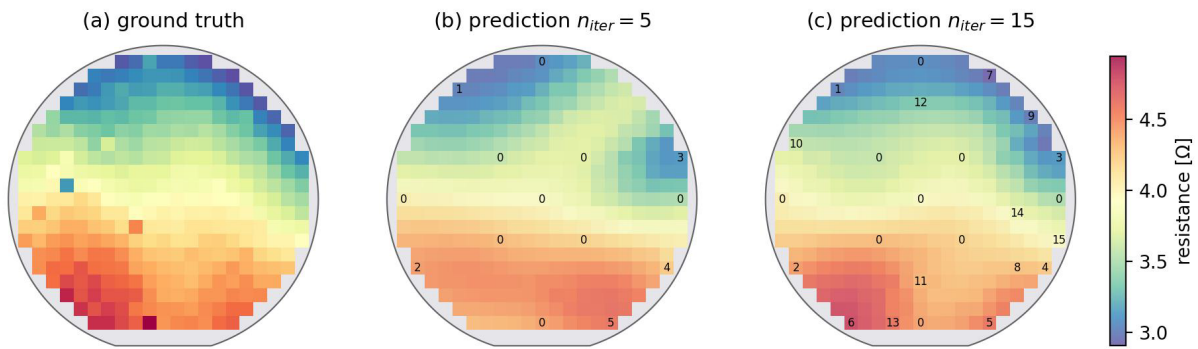

**Figure 3:** Comparison of the measured resistance distribution of the Co-Cr-Fe-Mn-Mo library (a) to two different predictions after 5 (b) and after 15 iterations (c). After 15 iterations, the ground truth and prediction are already nearly identical. The compositional data and the areas coordinates were used as an input. All inputs and outputs were standardized.

*Choice of input parameters*
In order to give the active learning algorithm additional information of the library to be measured, the chemical composition determined by EDX (Oxford X-act, accuracy: 1 at. %) was used as an input for the algorithm. Since EDX is normally done directly after deposition of the library, the composition data is available prior to the resistance measurements. Depending on acceleration voltage and materials, the electron beam reaches different depths up to several micrometers. Therefore, not only the deposited elements, but also the substrate material can be included in the analysis. This can support autonomous resistance measurements, as substrate information is generally correlated with film thickness, which in turn influences the electrical resistance.

In order to test the influence of the selection of constituents, the performance of a standard Gaussian process with SE kernel was observed, trained either on the compositional information of the deposited elements only, or on data including the substrate contents. The (normalized) x- and y-coordinates were added to the training data as well, to give the Gaussian process positional information of the resistance distribution on the library. Input and output standardization was used to improve numerical stability. The results of the first 250 training iterations are shown in Figure 4. For even more iterations, the Gaussian process tends to memorize the added training data, generally referred to as overfitting. Therefore, following iterations are neglected. Across all tested libraries, an accuracy higher than 90% after 50 iterations was observed using the standard implementation of the Gaussian process. The highest performance was achieved for the measurement of libraries which generally show unidirectional resistance gradients (the first five in Table 1).

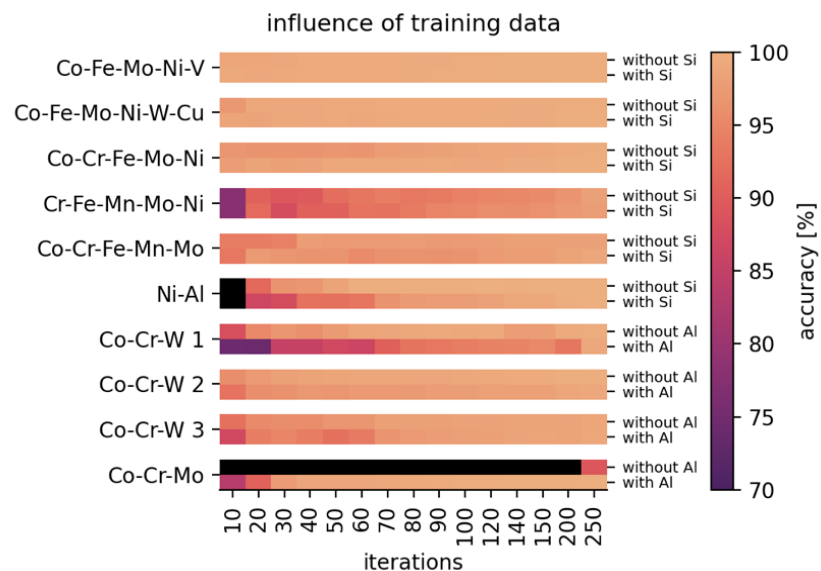

**Figure 4:** Comparison of the performance of the active learning optimization when including the substrate content (either Si or Al) into the training data. When removing a constituent from the EDX data, the compositions were renormalized to a content of 100 at.%. A standard Gaussian process was used, input and output training data were standardized, and the coordinates of the measurement areas were added to the training data. Before the optimization loop, the Gaussian process was trained on 9 initialization measurement areas (see Figures 2 and 3).

Including the substrate information in the training data was mostly found to either not affect the performance or slightly improve the accuracy and robustness of the prediction. This is because the electrical resistance depends both on composition and thickness. Only in one case (Co-Cr-Mo) of the tested 10 different libraries the inclusion of substrate information showed a substantially improved result. For the libraries Co-Cr-W 1 and Ni-Al the Gaussian process shows a decrease in performance when trained on the substrate information, indicating that the resistance mainly depends on the material rather than the thickness. Additional noise brought into the training data by the substrate information is not visibly affecting the performance. Consequently, as including thickness information via the substrate content was shown to be generally enabling a more robust prediction, as much information as available should be added to the training data.

*Incorporation of the measurement standard deviation into the model*

There are generally two types of measurement errors which can occur during a four-point probe measurement. Due to the spring mechanism of the sensor pins, the contact to the sample can be slightly different between touchdowns, resulting in resistance deviations of 0.1 - 2 %. Up to 1-2 times during an entire library mapping, the contact pins can touch each other, resulting in a short circuit which lets the source meter output values in the positive or negative MΩ-range. To account for these errors, every measurement area is measured three times, and during each contact ten resistance values are recorded. A standard Gaussian process is unable to work with ambiguous data in which multiple output data points are assigned to the same input data, which is why the mean of the conducted measurements is normally calculated prior to training. However, with this approach, available information about the reliability of the measurement results is lost. The solution is to modify the marginal likelihood of the Gaussian process. In a standard Gaussian process, the hyperparameters are estimated by maximizing the marginal likelihood given by

$$\log p(\boldsymbol{y}|\boldsymbol{X}) = \mathcal{N}(\boldsymbol{y}|0, \boldsymbol{K} + \sigma_n^2 \boldsymbol{I}) \tag{1}$$

where $\boldsymbol{X}$ and $\boldsymbol{y}$ are the input and output training data, $\boldsymbol{K}$ denotes the covariance matrix and $\sigma_n^2$ the noise variance. A mean of $\mu = 0$ is assumed [19]. Instead of determining the noise variance via hyperparameter optimization, the variances $\boldsymbol{\sigma}_m^2$ of the output training data points obtained by the 30 individual measurements can be used to compute the marginal likelihood of the model [33].

$$\log p(\boldsymbol{y}|\boldsymbol{X}) = \mathcal{N}(\boldsymbol{y}|0, \boldsymbol{K} + \boldsymbol{\sigma}_m^2) \tag{2}$$

This enables the Gaussian process to automatically weigh the measurement results based on their reliability without modifying the standard Gaussian processes architecture significantly.

Figure 5 (a) compares the standard Gaussian process to the one trained with the measurement variance over the first 250 iterations. A full visualization can be found in the supporting information. Without outliers, both implementations show almost identical results, the mean deviation of accuracy across all tested libraries is 0.2%. This small improvement originates from the ability of the algorithm to detect minor measurement errors caused by variations of the pin's orientation during each individual measurement. In order to investigate the performance with higher measurement noise, the accidental short-circuit of the pins was simulated by adding randomly generated noise in the range of $0.8 - 1.2 \ M\Omega$ to three measurement areas across all libraries. The resulting resistance distributions can be found in the supporting information. In this simulation, it is assumed that one out of three touchdowns feature ten resistance measurement results with a large variance. The resulting performance of the vanilla Gaussian process and the one based on the measurement variance is shown in Figure 5 (b). While the standard Gaussian process fails predicting the distribution as soon as an outlier is measured, the active learning algorithm with integrated measurement variance continues the optimization once an outlier is reached, as it is capable of automatically weighting its output training data relative to its reliability.

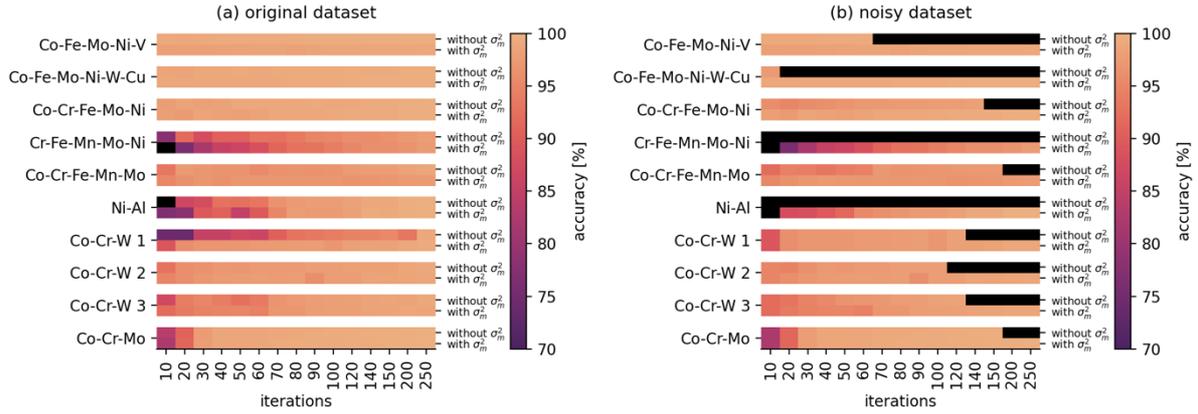

**Figure 5**: Performance of the standard Gaussian process and the one with information about the resistance measurement variance $\sigma_m^2$ on 10 materials libraries. For data with low noise levels (a), both algorithms show comparable performances with a mean deviation of 0.1%. When trained on a noisy dataset with three artificially added random outliers, the prediction of the standard Gaussian process fails as soon as a measurement area with an outlier is reached. The incorporation of the measurement variance required the deactivation of output standardization. A detailed visualization of both tests can be found in the supporting information.

*Kernel test*

As the kernel of a Gaussian process controls the shape of the regression function, choosing an appropriate kernel is important for ensuring a robust operation of the autonomous measurement. Here, four kernels are compared, the SE kernel, two kernels of the Matérn class, and the rational quadratic (RQ) kernel. Since the algorithm needs to be suitable for a large variety of different materials and libraries, sufficient adaptability and stability are the most important factors for choosing the kernel. Each library was autonomously measured with each kernel and ranked by their performance. The results are summarized in Table 2. The accuracy improvement over the iterations can be found in the supporting information. Except for the RQ kernel, the performances of the different kernels across all materials libraries were found being very similar. While there was no kernel performing best for each of the ten libraries tested, the Matérn kernels were found to have slightly better prediction accuracy. Reasons for this are their larger set of hyperparameters and the resulting greater flexibility. The rational quadratic kernel on the other hand was the only kernel unable to approximate all libraries and failed in four occasions entirely. For future uses of the algorithm, the kernel Matérn52 was chosen.

Table 2: Performance of different kernels on all test-libraries and the number of iterations dictated by the tested stopping criterion. For each library, the performances of the kernels were ranked (from 1 (worst) to 4 (best), a zero was given in case the algorithm failed the prediction entirely). The Matérn52 kernel performed best for most tests. In the third column, the number of iterations proposed by the stopping criterion are compared to optimal ones determined by observing a visual representation of the prediction as well the accuracy. For most cases, the stopping criterion is measuring more areas than needed.

| Material system | Kernels | | | | Stopping criterion | |
|---|---|---|---|---|---|---|
| | SE | RQ | Matérn32 | Matérn52 | n iterations | optimal |
| Co-Fe-Mo-Ni-V | 3 | 0 | 3 | 4 | 41 | 20 |
| Co-Fe-Mo-Ni-W-Cu | 3 | 2 | 1 | 4 | 41 | 10 |
| Co-Cr-Fe-Mo-Ni | 2 | 0 | 3 | 4 | 75 | 16 |
| Cr-Fe-Mn-Mo-Ni | 2 | 0 | 3 | 4 | 41 | 35 |
| Co-Cr-Fe-Mn-Mo | 2 | 1 | 4 | 3 | 48 | 40 |
| Ni-Al | 3 | 3 | 4 | 2 | 63 | 80 |
| Co-Cr-W 1 | 3 | 4 | 2 | 2 | 41 | 26 |
| Co-Cr-W 2 | 3 | 3 | 4 | 4 | 50 | 50 |
| Co-Cr-W 3 | 3 | 2 | 2 | 4 | 41 | 38 |
| Co-Cr-Mo | 3 | 3 | 2 | 4 | 59 | 47 |
| Mean | 2.7 | 1.8 | 2.8 | 3.5 | | |

*Stopping criteria*

By formulating a suitable stopping criterion, the measurement process can be terminated before all areas are measured in order to increase the efficiency of the measurement procedure. A robust implementation is the most important factor when choosing a stopping criterion, as it needs to be applicable for a wide variety of libraries and directly influences the final accuracy of the measurement. Outside the test environment, the accuracy (e.g., quantified by the coefficient of determination) of the algorithm cannot be used as a stopping criterion, since the ground truth is unknown prior to the actual measurement. Therefore, independent stopping criteria need to be considered. A static approach is to stop the autonomous measurement after a specific number of iterations determined by testing of a variety of different libraries. However, given the large variety of materials to characterize with the resistance measurement, it is unlikely to find a quantity of iterations suitable for all experiments. Another approach that is easy to implement is a human-in-the-loop [10], who can stop the measurement process by interacting with a graphical user interface. This supervisor can then judge the quality of the optimization based on the current state, which requires the supervisor's attention and availability at all times.

In order to overcome this, a dynamic stopping criterion based on the predicted uncertainty of the Gaussian process is proposed. However, simply defining an uncertainty threshold under which the process is terminated is not applicable either, as each measured library will have a different range of uncertainties depending on the noise level of the measurement and potential outliers. Therefore, the uncertainty over the training iterations needs to be observed relative to the initial uncertainty. The stopping logic is shown in Figure 6 on the example of the Co-Cr-Fe-Mo-Ni library. The (unknown) accuracy of the optimization process, the normalized mean covariance predicted by the Gaussian process as well as the numerically determined gradient of the normalized mean covariance are plotted over the training iterations.

After initialization of the Gaussian process, 30 areas are measured independent of the performance ensuring a basic approximation of the dataset. Afterwards, the normalized predicted covariance of each iteration is observed. If the covariance of the current iteration is smaller than the initial covariance, the numeric derivative of the normalized mean covariance is calculated, and its progression is observed over the next ten iterations. This criterion is driven by the notion, that in order to terminate the process, the model at least needs to have an uncertainty lower than the one of the initial iteration. If the model continues to improve its fit over the next ten iterations (indicated by a steady decrease in the mean covariance), the measurement process is stopped. This is determined by observing the gradient of the mean covariance, specifically by ensuring that it stays below the empirically found threshold of $1\ \%/iter$. Otherwise, if the gradient is positive, the observation is reset and at least ten additional measurements are taken until the next termination is possible.

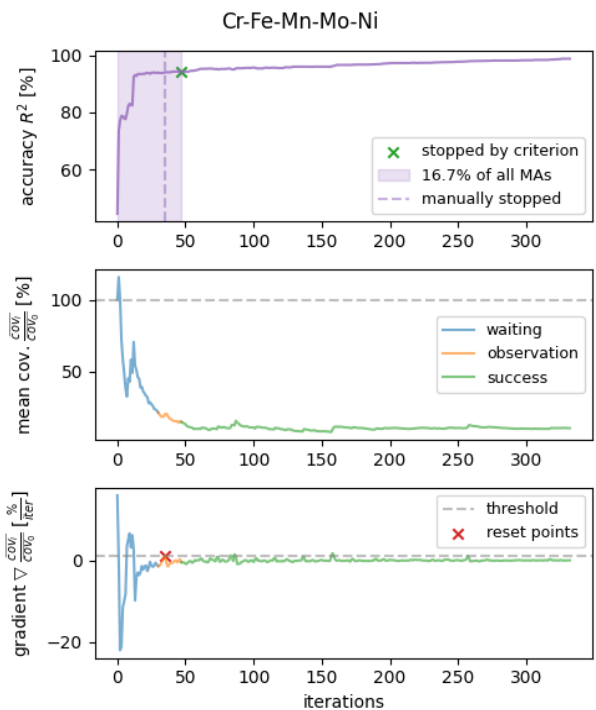

**Figure 6:** The developed dynamic stopping criterion on the example of the autonomous measurement of the Cr-Fe-Mn-Mo-Ni library. First, thirty measurements are done independent on the performance of the algorithm. Afterwards, the normalized mean covariance is observed over the following iterations. As soon as the current uncertainty becomes smaller than the one of the first prediction, the change of the gradient over the next ten iterations is observed. If this numerically determined gradient of the normalized mean covariance is staying below the threshold of $1\ \%/iter$ the process is stopped. A second increase of the mean covariance is indicated by the reset points. In case of the Cr-Fe-Mn-Mo-Ni library, the process is stopped after 47 iterations which corresponds to 16.7% of all measurement areas.

The stopping criterion applied to the other libraries can be found in the supporting information. Table 2 compares the number of iterations determined via the shown stopping criterion and the optimal stopping decision based on observing the accuracy of the algorithm as well as a visual representation of the prediction. In most cases, the developed stopping criterion is overestimating the number of measurements to perform by a factor of 1.5-4. Although this can be finetuned by changing the fixed number of initial iterations or the number of iterations in which the mean covariance is supposed to decrease, this behavior is beneficial for this early implementation of the algorithm. In order to apply autonomous measurements to real-world every-day scientific workflow, enough trust in this technology needs to be established, that is why higher safety margins are useful during early adoption. However, for most tested libraries, the autonomous measurement could ideally be stopped after 6-16 % of the normally measured areas of a library without loss in quality. This applies especially to the tested co-sputtered HEA libraries, which feature uniform resistance gradients with less resistance variations. The Ni-Al library and two Co-Cr-W libraries are not predicted without any loss in quality even after measuring the entire library. Reason for this is missing information in the training data, as the resistance distribution changes are not fully reflected in the compositional data as well as an overall more inhomogeneous resistance distribution. For further studies, the incorporation of visual or crystal structure information could help improving the prediction in those cases.

**Conclusion**

The presented active learning approach for autonomous measurements shows great potential in increasing the efficiency of combinatorial experiments. Depending on the measured materials library, a measurement time reduction of about 70-90% was observed when considering the optimal iteration for stopping the process. As there is no criterion resulting in the optimal stopping for every experiment, the number of measurements to perform are increased by a factor of 1.4 - 4 when using the developed dynamic stopping criterion. The autonomous measurement procedure was implemented into the existing measurement device and can be used on the daily basis.

In order to gain additional insight and trust into the autonomous measurement procedure, the performance of the method can be evaluated in a longtime study while still measuring the entire library and therefore not taking the risk of less accurate or even wrong experiment results. Despite the achieved high efficiency improvement, the autonomous measurement only decreases the absolute measurement duration by about 30-40 minutes due to the already fast four-point probe measurement procedure. Therefore, the application into materials characterization devices demanding much more time can result in even higher absolute efficiency improvements. An example is the area of temperature-dependent resistance measurements, as temperature cycling is inherently slow with 20-50 hours [34] depending on the number of temperature steps and the temperature interval. Further, the widely used EDX or XRD measurement techniques can profit from active learning optimization as well. Further progress in these areas depends on manufacturers, who need to provide APIs for their highly specialized devices, which would enable intervening into the measurement processes via custom made software.

**Author contributions**

Felix Thelen: conceptualization, data curation, formal analysis, investigation, materials library fabrication, software, validation, visualization, writing – original draft. Lars Banko: conceptualization, validation, supervision, writing – review & editing. Rico Zehl: materials library fabrication, investigation. Sabrina Baha: materials library fabrication. Alfred Ludwig: conceptualization, resources, writing – review and editing, supervision, project administration.

**Conflicts of interest**

There are no conflicts to declare.

**Acknowledgements**

This work was partially supported from different projects. A. Ludwig and F. Thelen acknowledges funding from the Mercator Research Center Ex-2021-0034 and from the Zentrales Innovationsprogramm Mittelstand KK5380601ZG1 (ZIM, Central Innovation Program for Small and Medium-sized Enterprises). Funding by the Deutsche Forschungsgemeinschaft (DFG, German Research Foundation) Project 190389738 are acknowledged by R. Zehl and DFG Project LU1175/31-1 by S. Baha. For their support in synthesizing the material libraries, the authors would like to thank the student research assistants Katherine Guzey, Cedric Benedict Kaiser and Annika Gatzki. ZGH at Ruhr-University Bochum is acknowledged for the use of the scanning electron microscope.

**Data availability statement**

The source code as well as the composition and resistance data of this study are publicly available at [pub.matinf.pro/rubric/active-learning-resistance](pub.matinf.pro/rubric/active-learning-resistance).

**Supporting Information**

**Figure S-1:** Visualization of each of the ten test libraries containing a photo of the library, compositional data, and the measured electrical resistance.

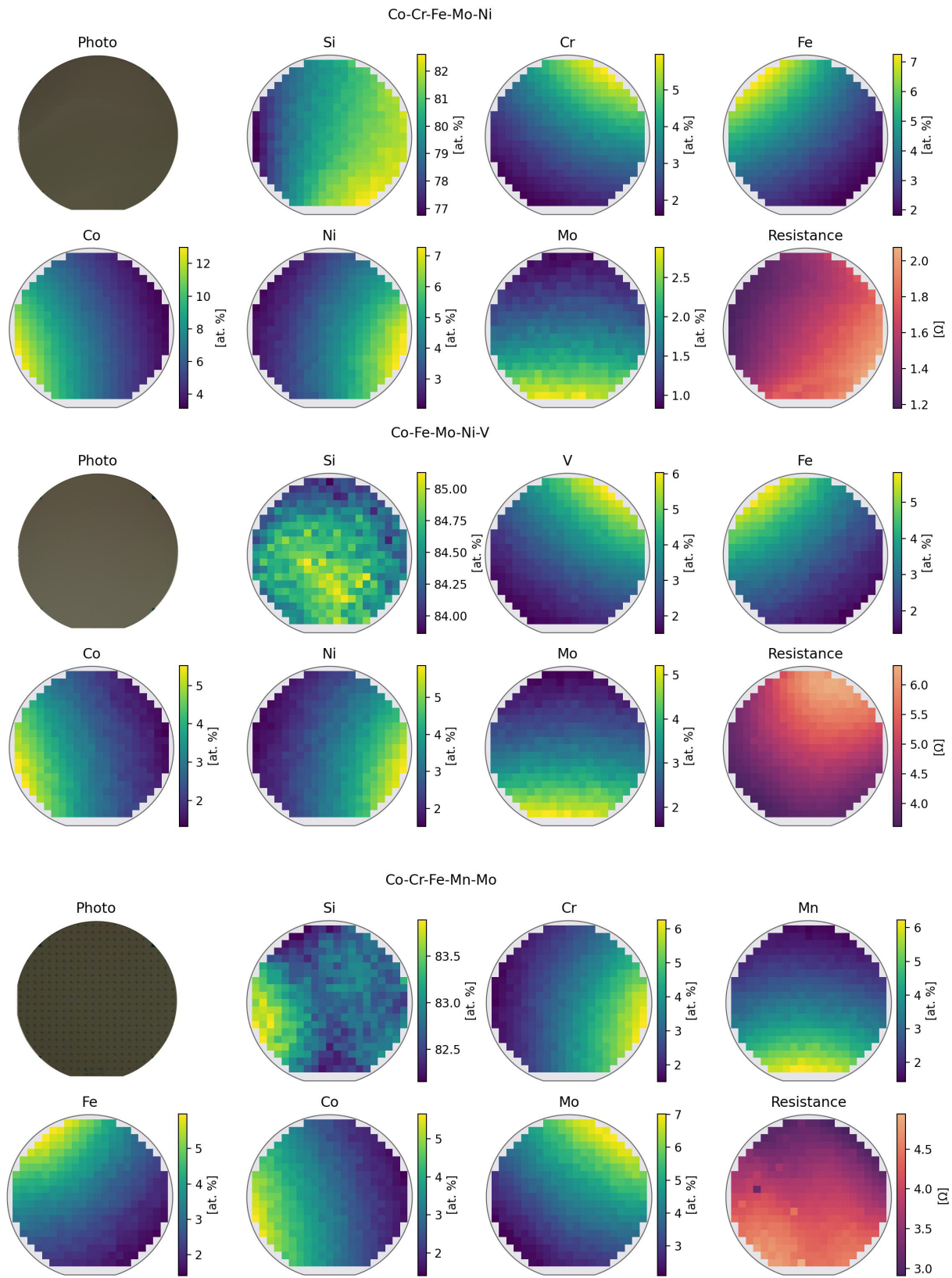

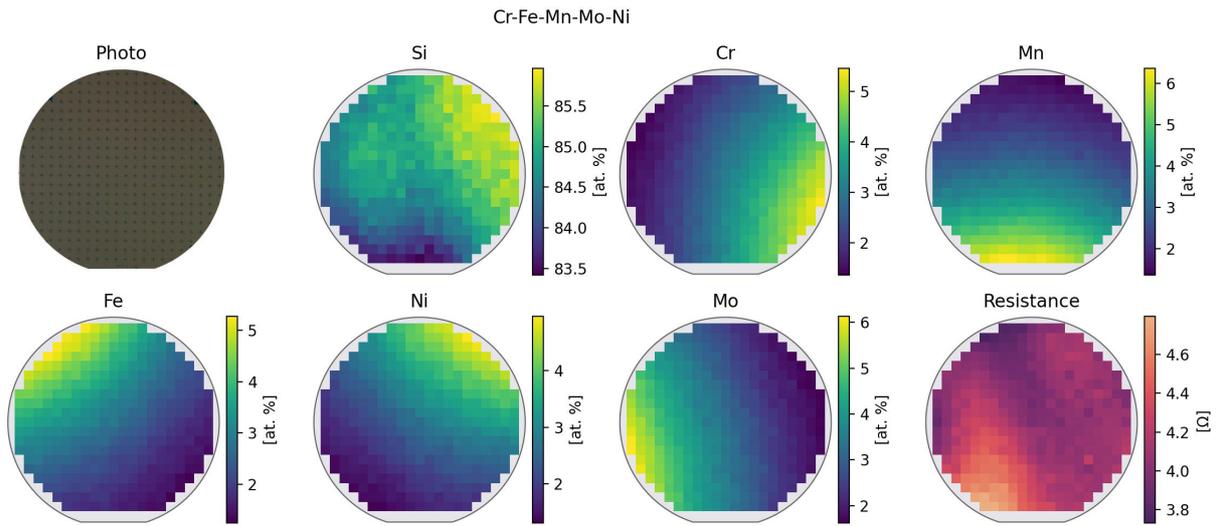
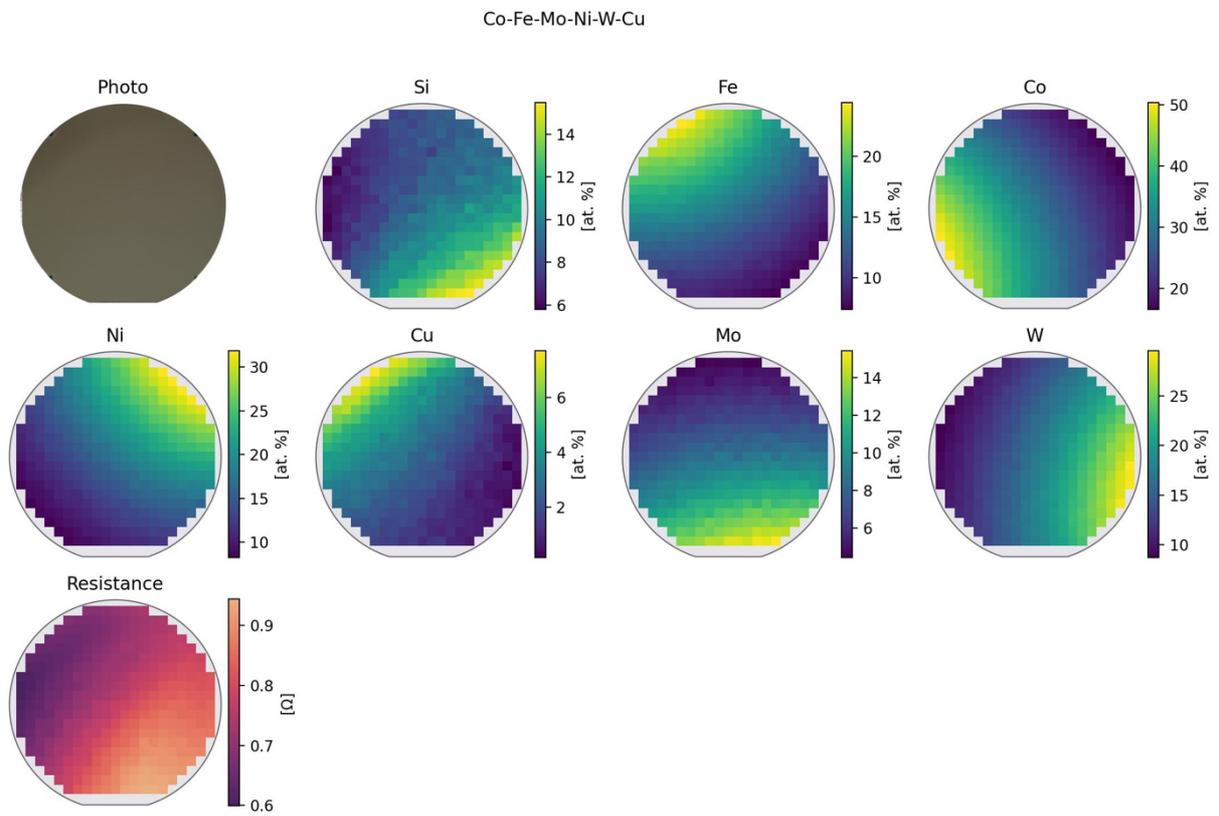

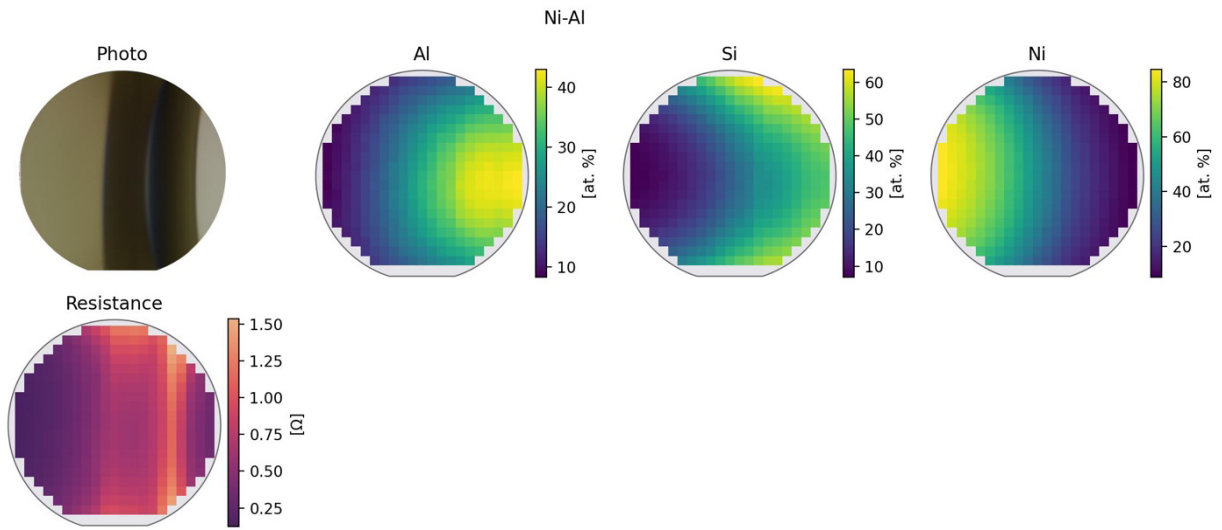
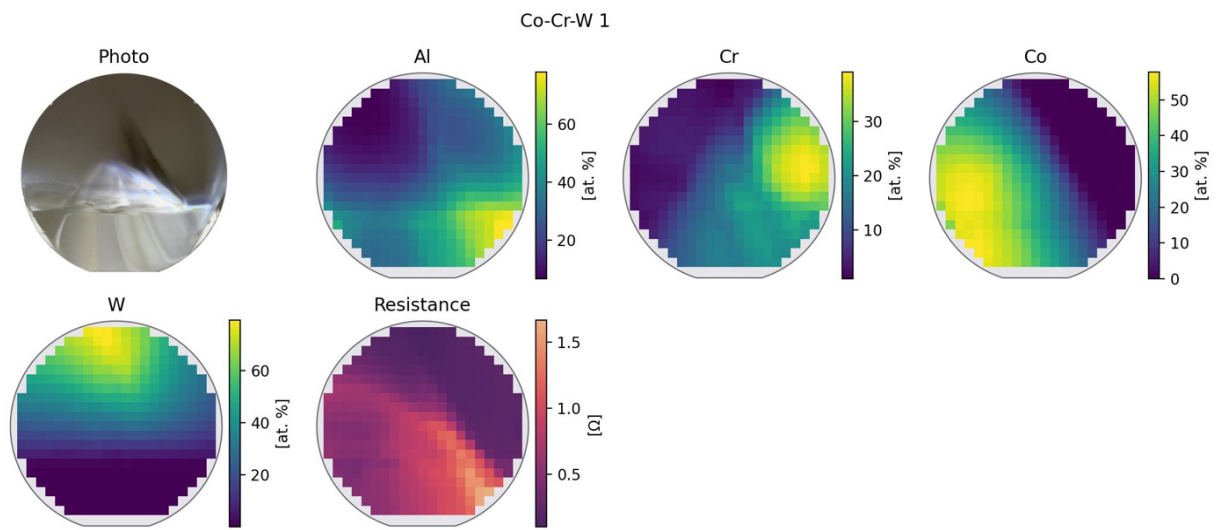
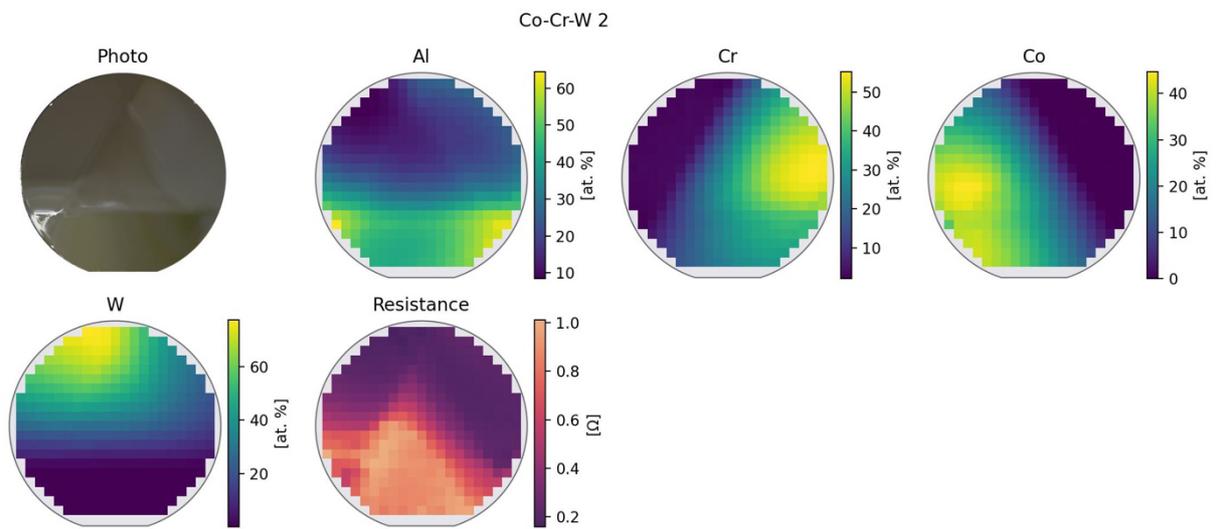

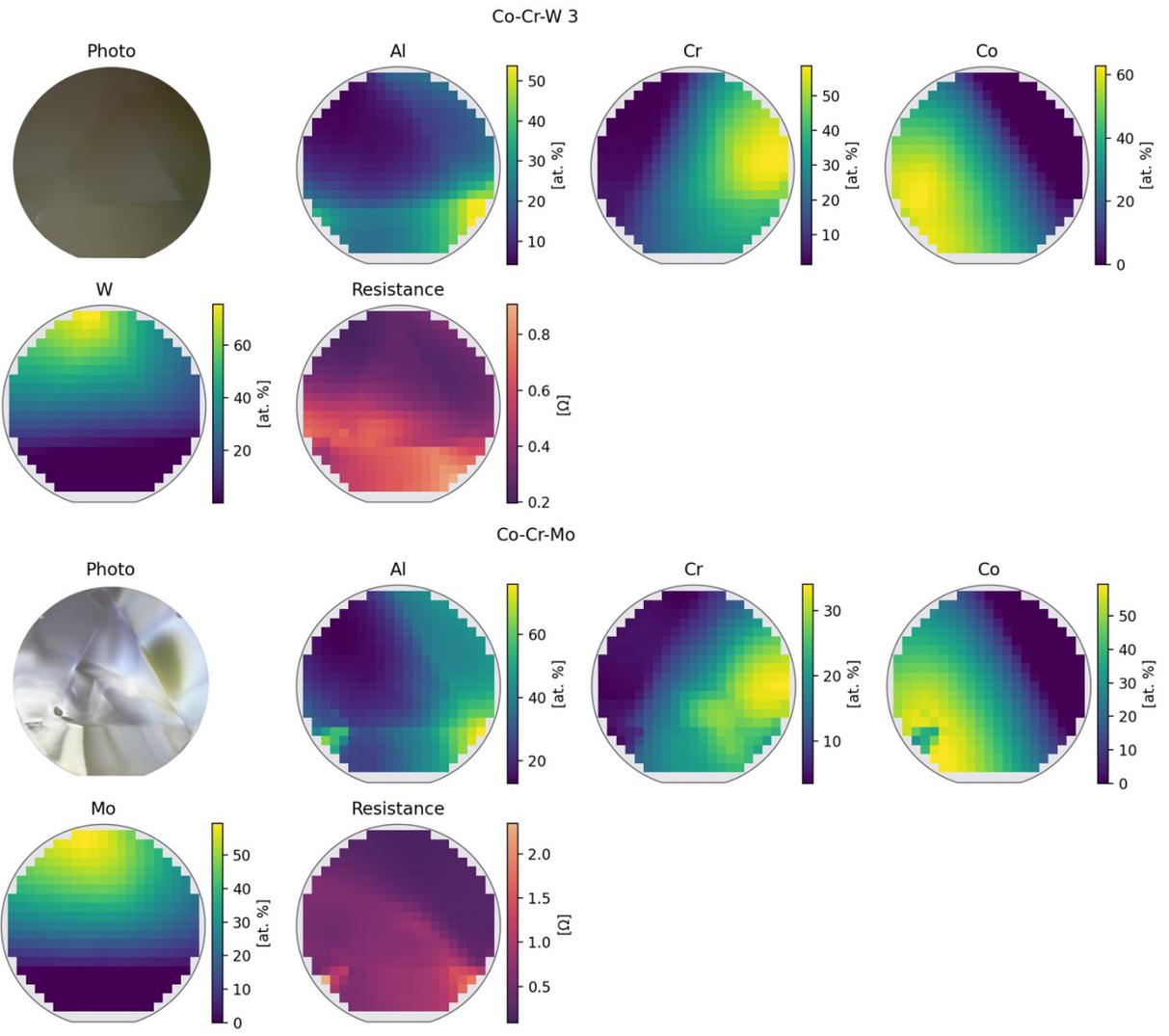

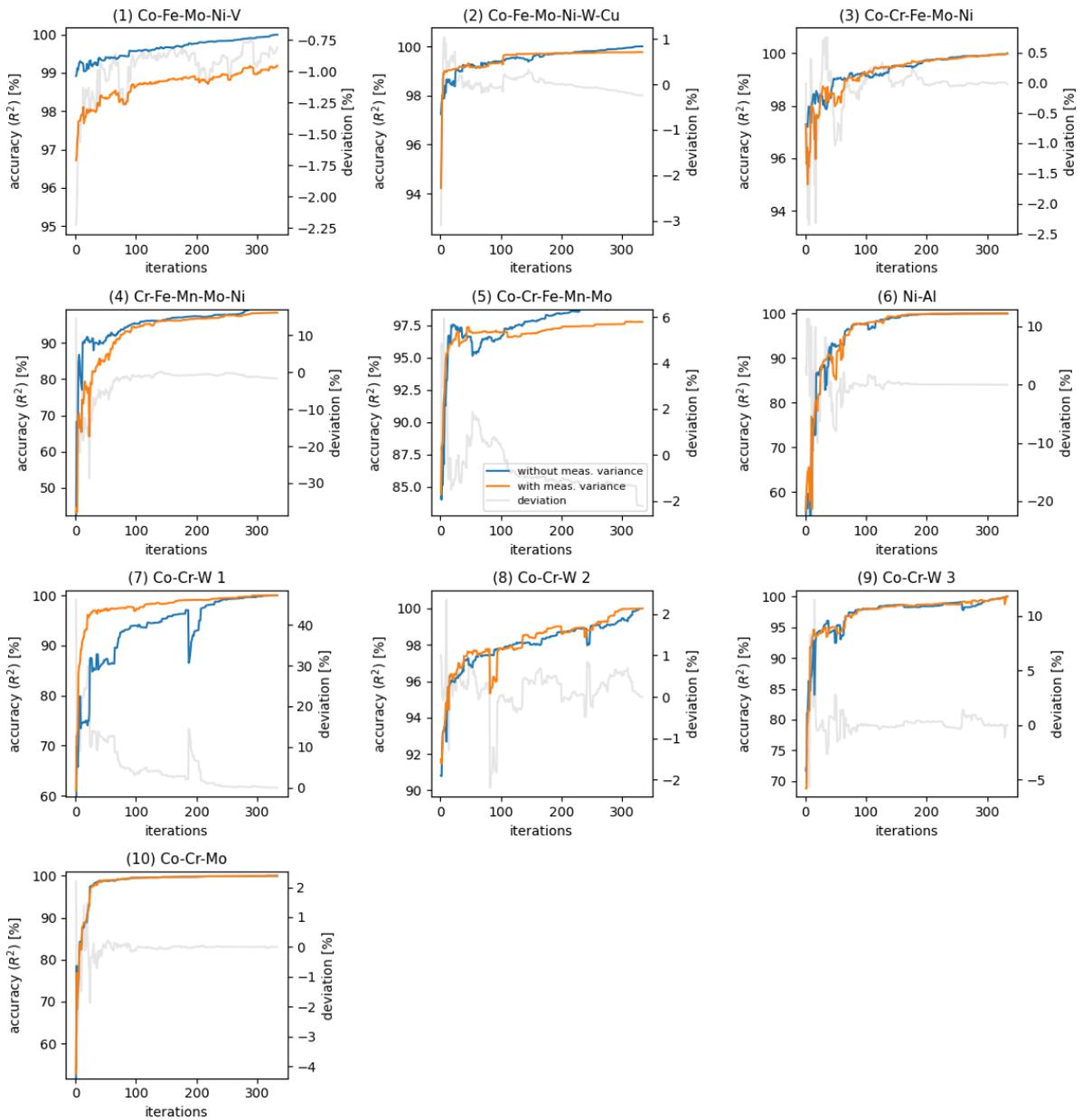

**Figure S-2:** Comparison of the GP performance on the dataset shown in Figure S-1 with and without including the measurement variance.

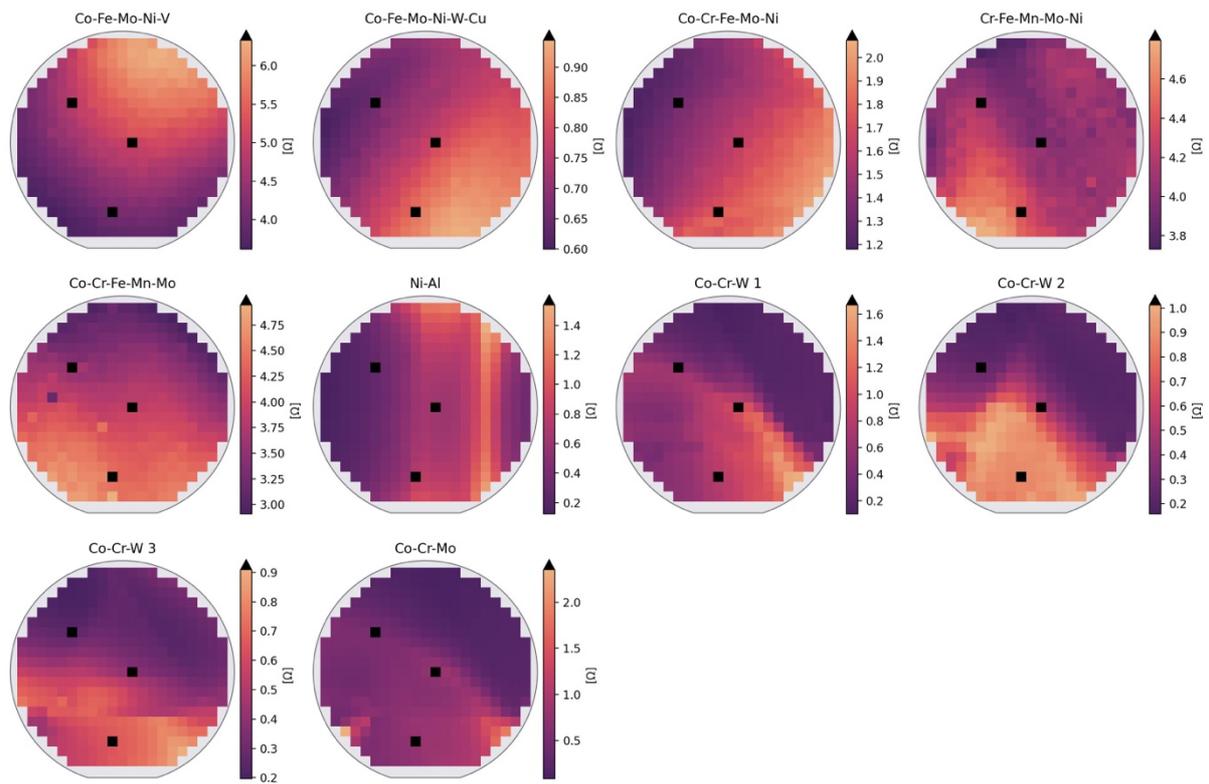

**Figure S-3:** Dataset with simulated outliers. As a single failed touchdown of the pins should be simulated, ten out of 30 resistance measurements were exchanged by outliers chosen randomly between $0.8 - 1.2\ M\Omega$. The position of the outliers was fixed to ensure comparability. Each plot shows the mean resistance at each measurement area.

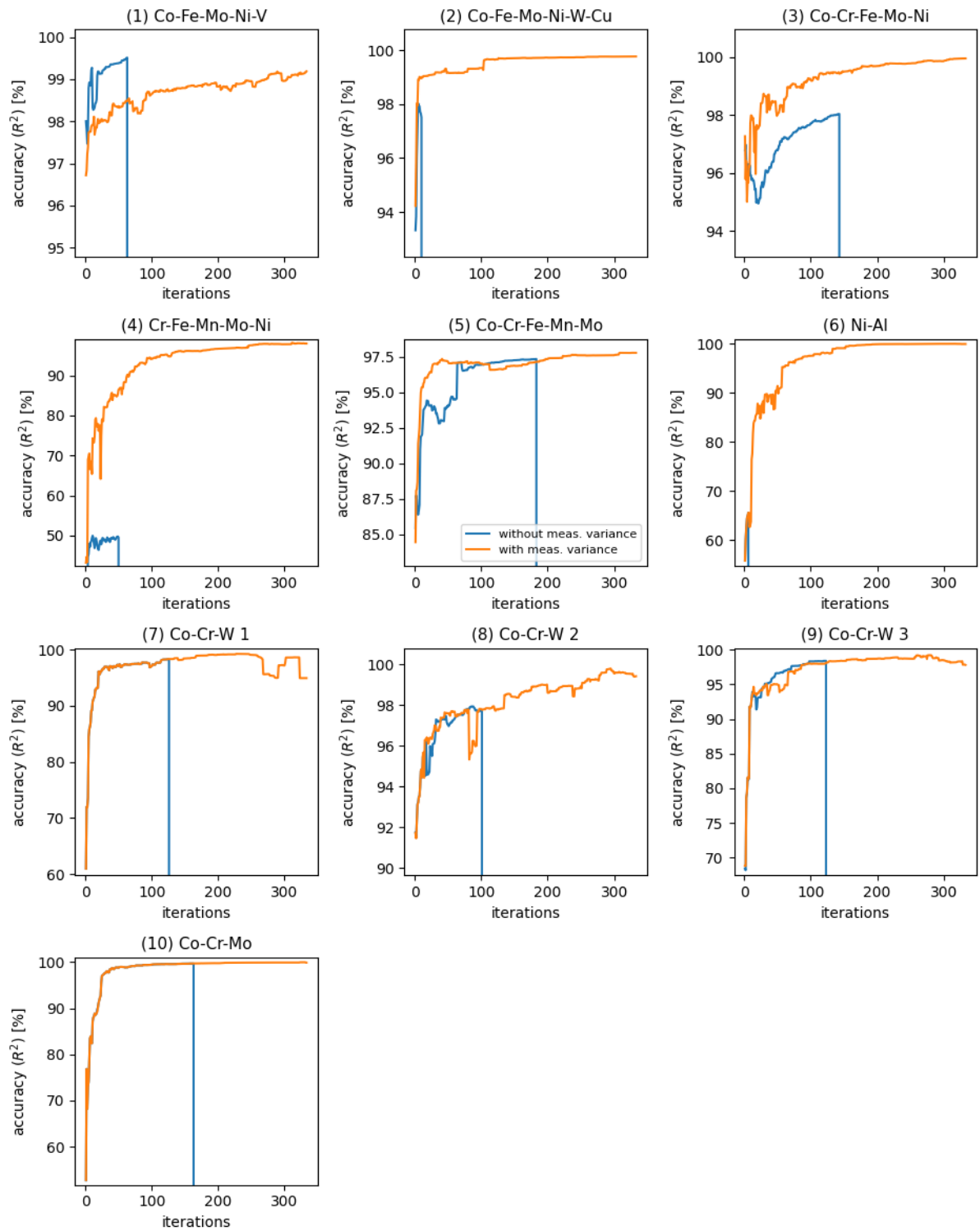

**Figure S-4:** Visualization of the resistance distributions with randomly added measurement noise. As soon as the vanilla GP encounters an outlier, the prediction fails, while with incorporating the measurement noise into the model, the GP is able to skip the outliers.

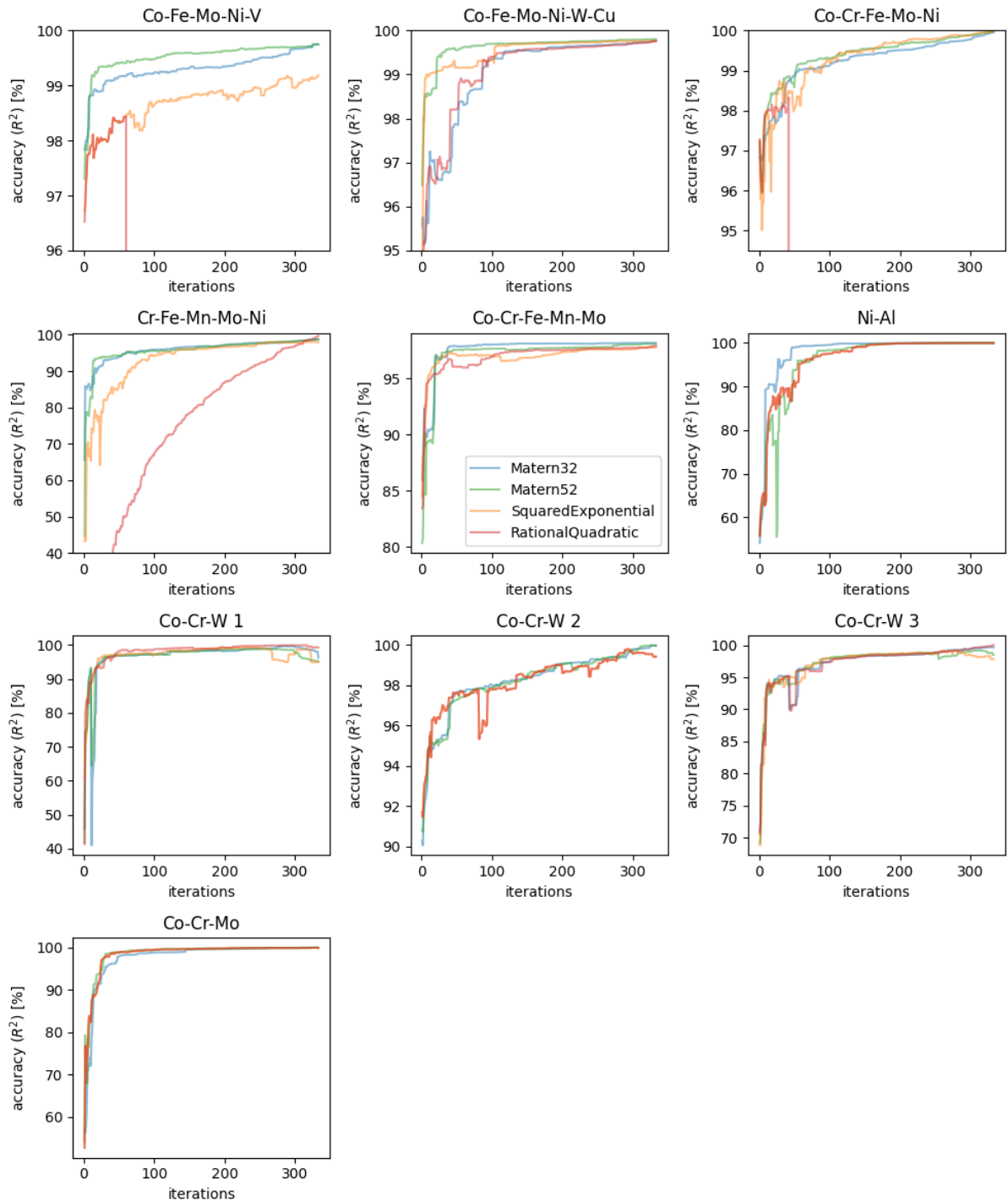

**Figure S-5:** Performance of the autonomous measurement with different Gaussian process kernels. Two kernels of the Matérn kernel class, the squared exponential as well as the rational quadratic kernel are compared.

**Figure S-6:** Visualization of the developed stopping criterion for all tested materials libraries. The accuracy of the optimization process, the mean covariance and well as the gradient of the mean covariance is shown over the iterations until all MAs are measured. The stopping iteration is marked in green, while the purple dashed lined shows the stopping iteration determined by observing the accuracy as well as a visual representation of the prediction (the optimal stopping opportunity). The purple range show the percentage of measured values compared to the entire library. The different colors of the mean covariance plot signal the different phases of the stopping criterion.

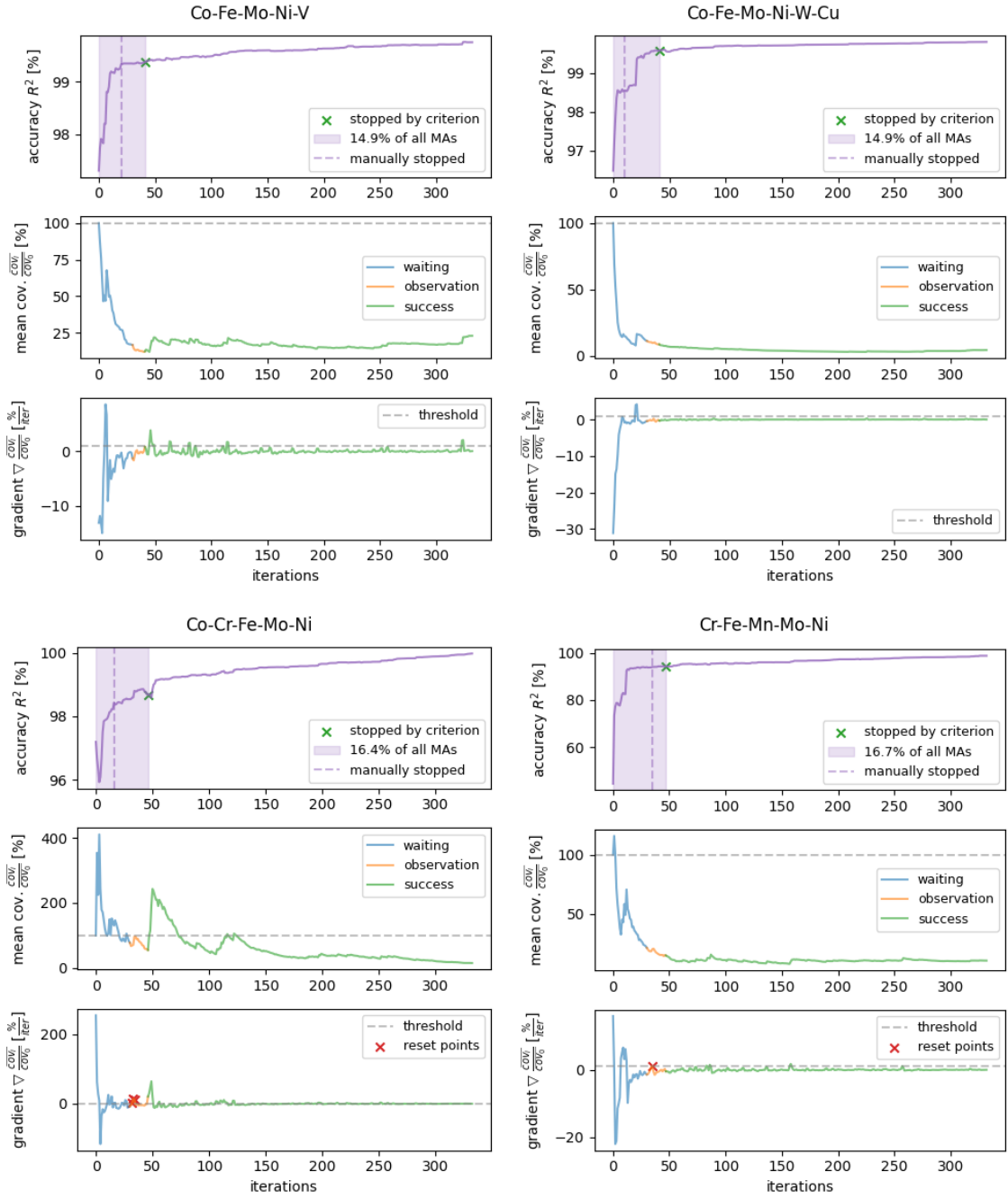

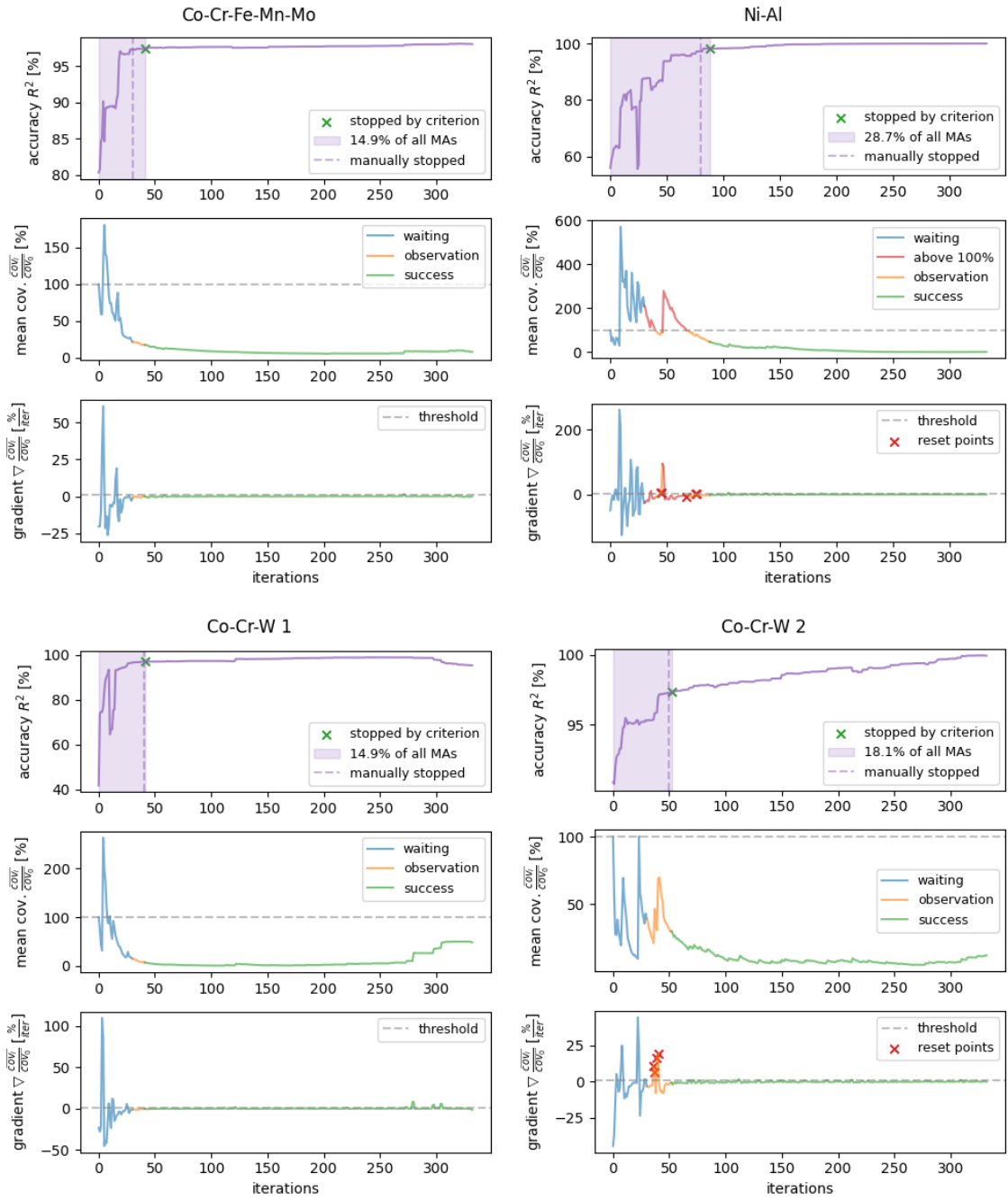

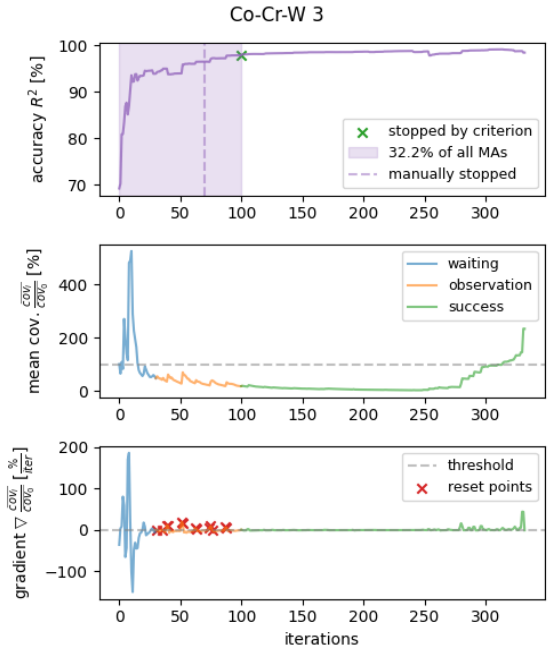
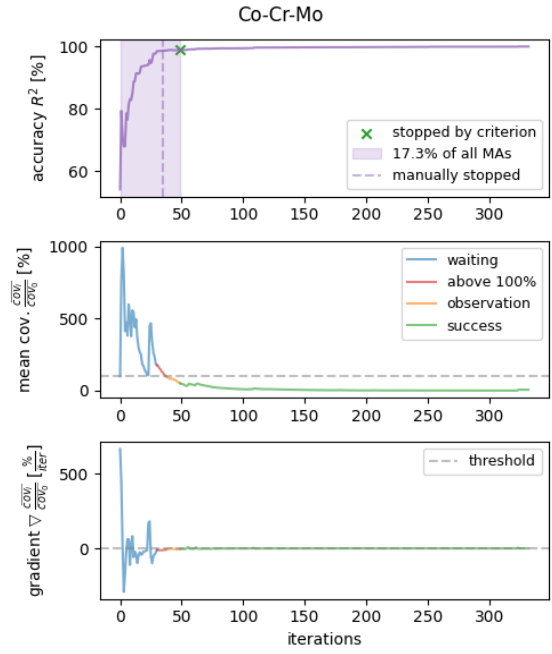